\documentclass[11pt]{article}
\usepackage{epsf,epsfig,here}
\usepackage{graphics}
\usepackage{graphicx}
\usepackage{epic}
\usepackage{floatflt}
\usepackage{dcolumn}
\usepackage{amsmath}
\RequirePackage{xspace}

\begin{document}
\vspace*{-1.8cm}
\begin{flushright}
{\bf LAL 06-173}\\
\vspace*{0.1cm}
{December 2006}
\end{flushright}

\vskip  1. cm
\begin{center}
{\LARGE\bf The BiPo detector for ultralow\\
\vspace*{0,2cm}
 radioactivity measurements}
\end{center}
\vskip 1.0truecm
\begin{center}
{\bf \large M. Bongrand}\\ 
{\small(on behalf of the SuperNEMO collaboration)}\\
\vspace*{0,4cm}
{\bf Laboratoire de l'Acc\'el\'erateur Lin\'eaire,}\\
{IN2P3-CNRS et Universit\'e de Paris-Sud 11, BP 34, 
F-91898 Orsay Cedex}\\
\end{center}

\vspace*{0.5cm}

\begin{abstract}
The BiPo project is dedicated to the measurement of extremely low radioactivity contamination of SuperNEMO source foils ($^{208}$Tl $<$~2~$\mu$Bq/kg and $^{214}$Bi~$<$~10~$\mu$Bq/kg). The R\&D phase is started: a modular BiPo prototype with its shielding test facility is under construction. The goal of this prototype is to study the background and particularly the surface contamination of scintillators. The first capsule has been installed in the Canfranc Underground Laboratory in October, 17$^{th}$ and is now taking data. After 10.7 days of measurements, a preliminary upper limit on the surface radiopurity of the scintillators of A($^{208}$Tl)~$<$~60~$\mu$Bq/m$^2$ (90\% C.~L.) has been obtained.
\end{abstract}

\section{Introduction}

 The BiPo detector is dedicated to the measurement of the ultra high radiopurity in $^{214}$Bi and $^{208}$Tl of ultra thin materials and especially the double beta source foils of the SuperNEMO detector. The expected sensitivity is $^{208}$Tl~$<$~2~$\mu$Bq/kg and $^{214}$Bi~$<$~10~$\mu$Bq/kg.

In order to measure $^{208}$Tl and $^{214}$Bi contaminations, the original idea of the BiPo detector is to detect the so-called Bi-Po process, a double detection of an electron followed by a delayed alpha, with organic scintillators. The $^{214}$Bi isotope is nearly a pure $\beta$ emitter (Q$_{\beta}$~=~3.27~MeV) decaying into $^{214}$Po, an $\alpha$ emitter with an half-life of 164~$\mu$s. The $^{208}$Tl isotope is measured by detecting its parent the $^{212}$Bi isotope. $^{212}$Bi decays with a branching ratio of 64\% via a $\beta$ emission in $^{212}$Po (Q$_{\beta}$~=~2.2~MeV) which is again an $\alpha$ emitter with a short half-life of 300~ns. So, for this two chains a BiPo signature is an electron and a delayed $\alpha$ with a delay time depending on the isotope contamination we want to measure.

\begin{figure}[h]
\centering
  \includegraphics[height=.15\textheight]{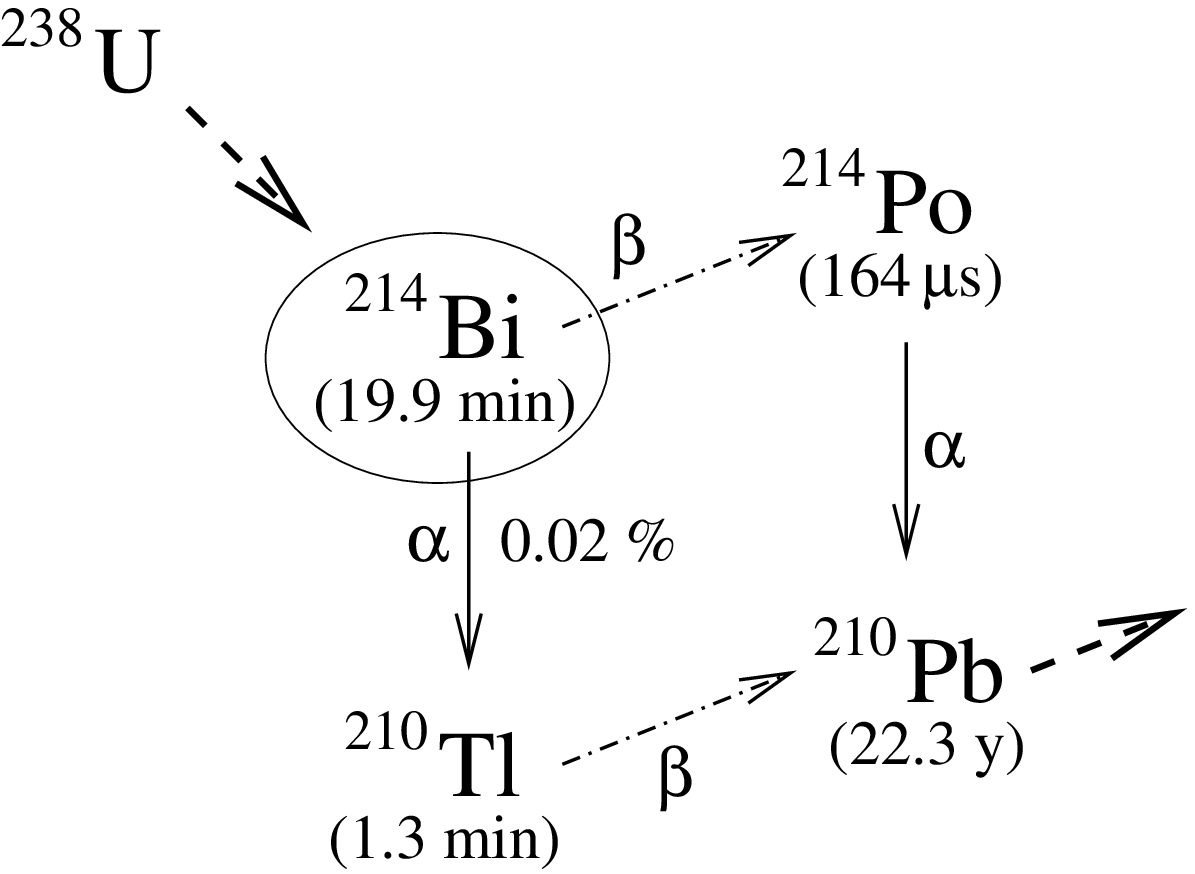}
  \hspace{1.5cm}
  \includegraphics[height=.145\textheight]{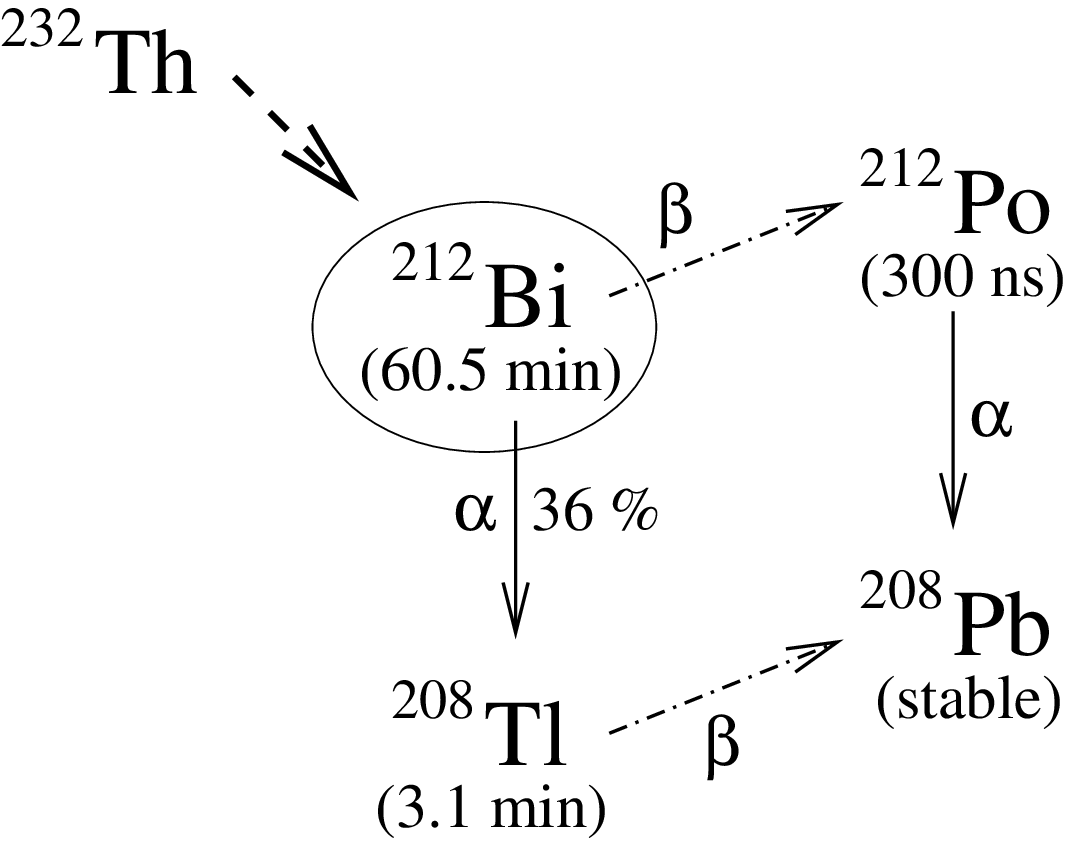}
  \caption{BiPo processes for $^{214}$Bi and $^{208}$Tl.}
  \label{fig:bipoproc}
\end{figure}

\section{Two possible designs}

Two designs for the BiPo detector are under study to find the best compromise between efficiency and background rejection.
\vspace*{0,2cm}
\subsection{Design with a tracking detector}

The first design, which derives the NEMO~3 technology, consists in a tracko-calo planar detector. The foil to be measured is deposited directly on a horizontal plane of organic plastic scintillator blocks (20$\times$20~cm$^2$) coupled to low radioactive 5'' PMTs. A drift wire chamber, working in Geiger mode, is installed above the foil. Long plastic scintillator bars, with two sides low activity PMTs readout, close the upper part of the wire chamber (figure \ref{fig:bipodesigns}). A BiPo event from a $^{212}$Bi decay is identified as a track of the electron in the wire chamber associated with a hit in time in a scintillator bar and a delayed hit within 1~$\mu$s ($\sim$3$\times$T$_{1/2}$~($^{212}$Po)) in a lower scintillator block near the reconstructed vertex. The geometrical efficiency is about 25\% since it requires the electron going up and the $\alpha$ going down.

The BiPo events from $^{214}$Bi decays is identified by a delayed $\alpha$, within 500~$\mu$s ($\sim$3$\times$T$_{1/2}$~($^{214}$Po)), in the tracking device. It is however difficult to tag this long delayed $\alpha$ by a delay hit in the lower scintillator block because of a probable too high random coincidence level within 500~$\mu$s.

The advantage of the design with a tracking detector is that it allows to measure $^{214}$Bi and $^{208}$Tl contaminations from natural radioactive chains with the same device. However, it requires to build a large gaseous detector and the geometrical efficiency is low.
\vspace*{0,2cm}
\subsection{Design without tracking detector}

\begin{figure}[h]
\centering
  \includegraphics[scale=0.35]{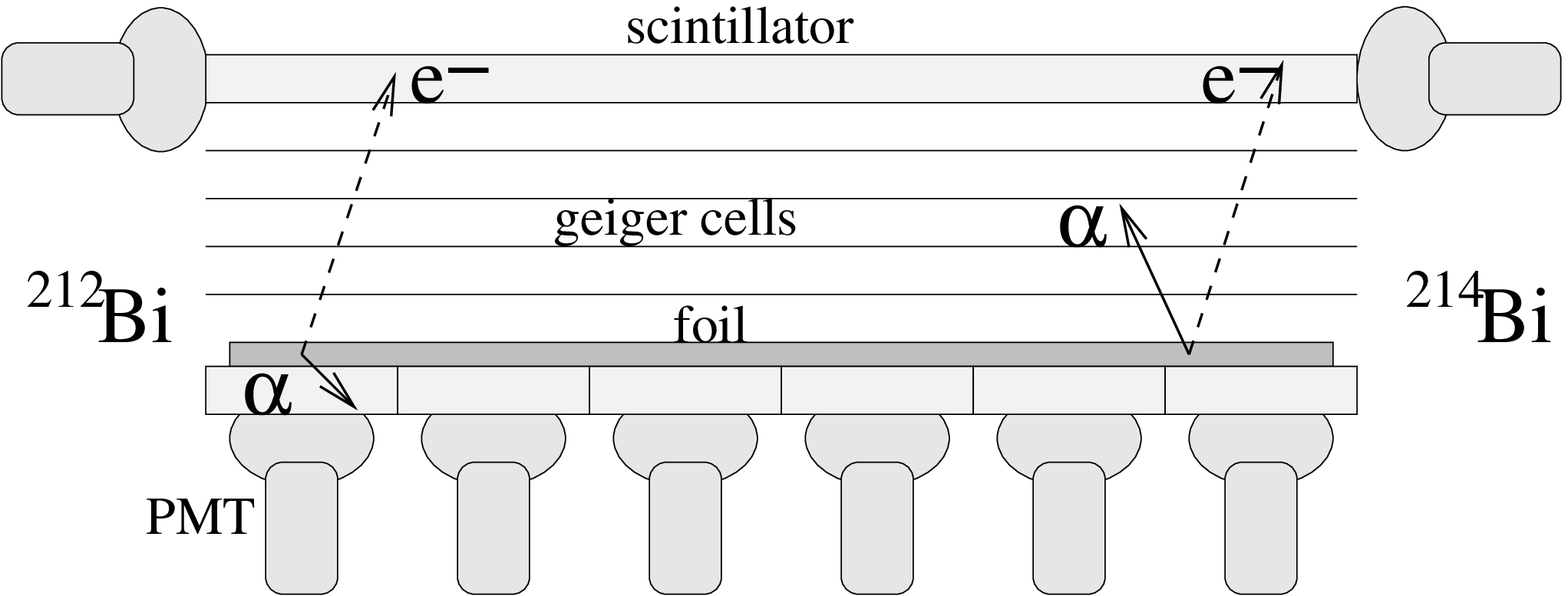}
  \hspace{1cm}
  \includegraphics[scale=0.27]{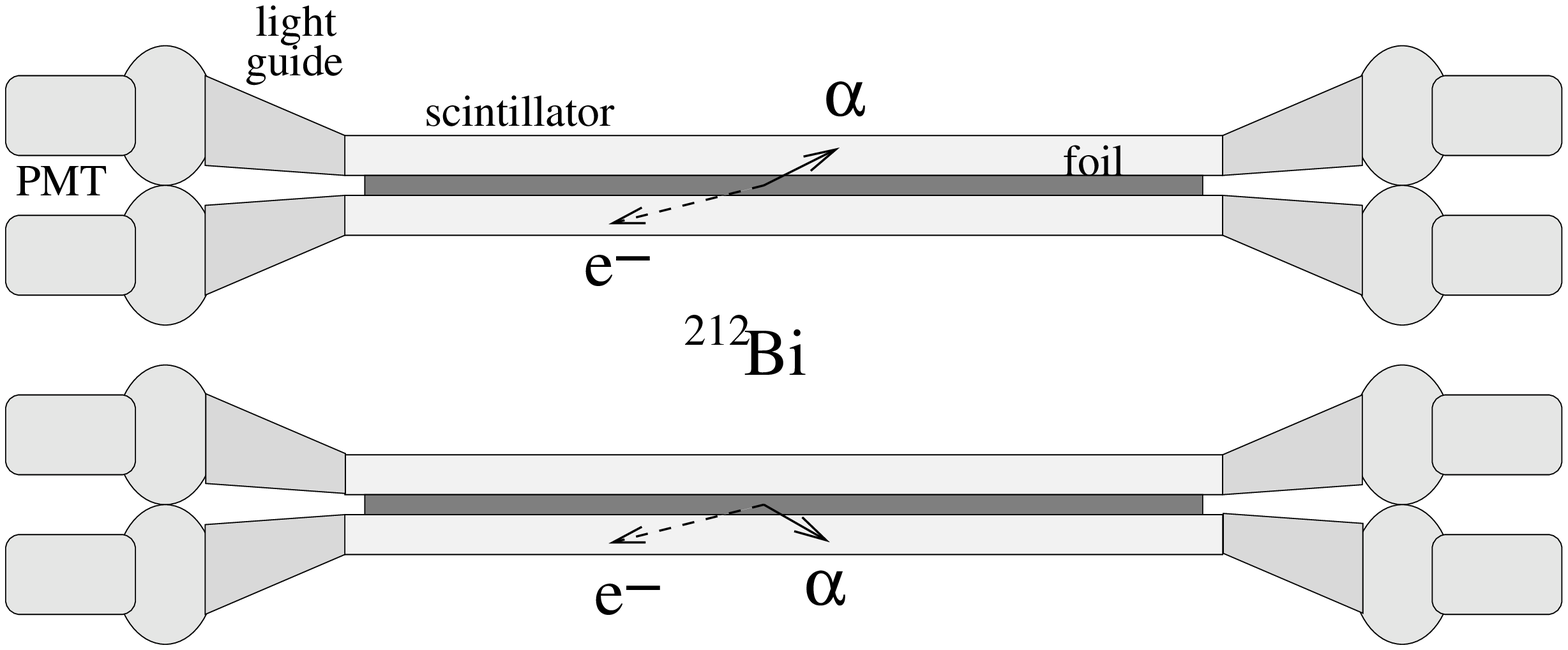}
  \caption{The 2 possible designs for the BiPo detector and the BiPo event signature: a prompt e$^-$ and a delayed $\alpha$.}
  \label{fig:bipodesigns}
\end{figure}

The second design consists of two thin organic plastic scintillating plates (1~cm thick) with the source foil sandwiched between this plates. The scintillating plates are polished without any wrapping in order to collect the scintillation light on the lateral sides by total internal reflectivity. The optical readout is done with low radioactive PMTs (figure \ref{fig:bipodesigns}). The position of the emission of scintillation light is reconstructed by the barycenter of the amount of light of each PMT. An external $\gamma$ tagger with plastic scintillators surrounds the detector in order to reject external background. A BiPo event corresponds to a trigger hit in one plate and a delayed hit from the same location in the same plate or in the second one and no hit in time in the $\gamma$ tagger. Let's notice that the $^{214}$Bi measurement will be difficult because of a probable too high random coincidences level within 500~$\mu$s.

This design has the advantage to be compact and to have a large geometrical efficiency.\\

\subsection{Ultra thin scintillating optical fibers}

The use of ultra thin scintillating optical fibers between the foil to be measured and the scintillators is a very attractive technique for electrons and alphas identification. A thickness of 200~$\mu$m allows both to contain totally and measure $\alpha$ particles and to tag electrons without loosing too much energy. This technique allows to reduce strongly the single rate (and so the random coincidence) and to suppress external $\gamma$ background. This technique may be adapted for both previous BiPo designs.

\section{Expected sensitivity of the BiPo detector}
\vspace*{0,2cm}
\subsection{$\alpha$ quenching factor}

The sensitivity of the BiPo detector is mostly correlated to the capacity of detecting the $\alpha$ emission. It requires a low energy threshold for $\alpha$ detection in order to be sensitive to the whole thickness of the source (40~mg/cm$^{2}$ in case of SuperNEMO foils). Moreover due to the very large stopping power of $\alpha$ particles, the amount of scintillation light produced by an $\alpha$ is smaller than the one produced by an electron of the same energy. A scintillation quenching factor $Q(E^{\alpha})$ for $\alpha$ particles depends on the energy of the $\alpha$ and is defined as:
\begin{equation}
E^{\alpha}_{meas} = \frac{E^{\alpha}}{Q(E^{\alpha})} \label{eqn:quenching}~.
\end{equation}

\begin{figure}[htb]
  \includegraphics[height=.1\textheight]{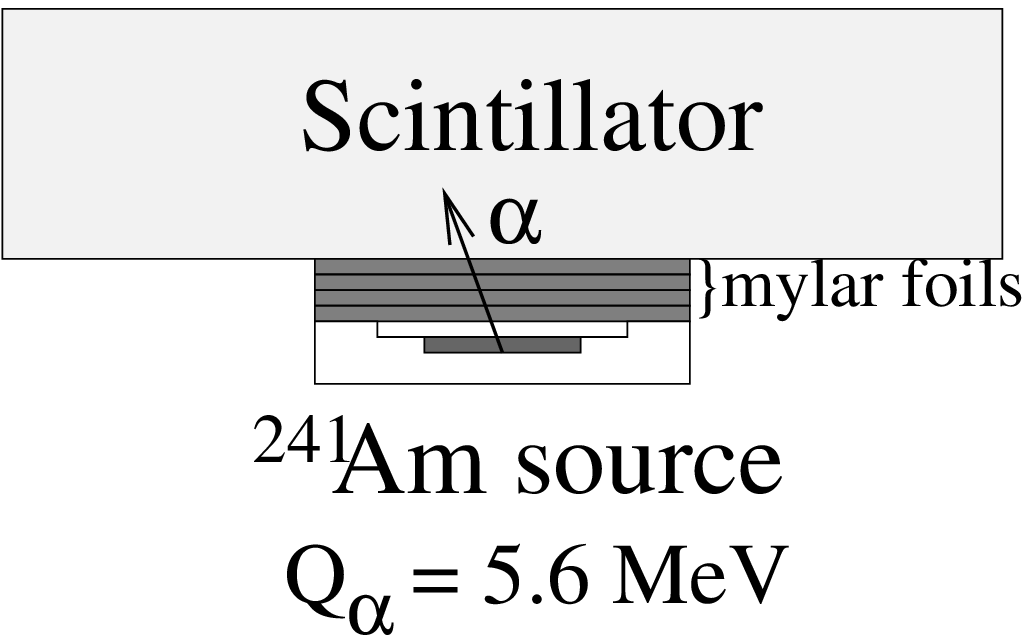}
  \includegraphics[height=.17\textheight]{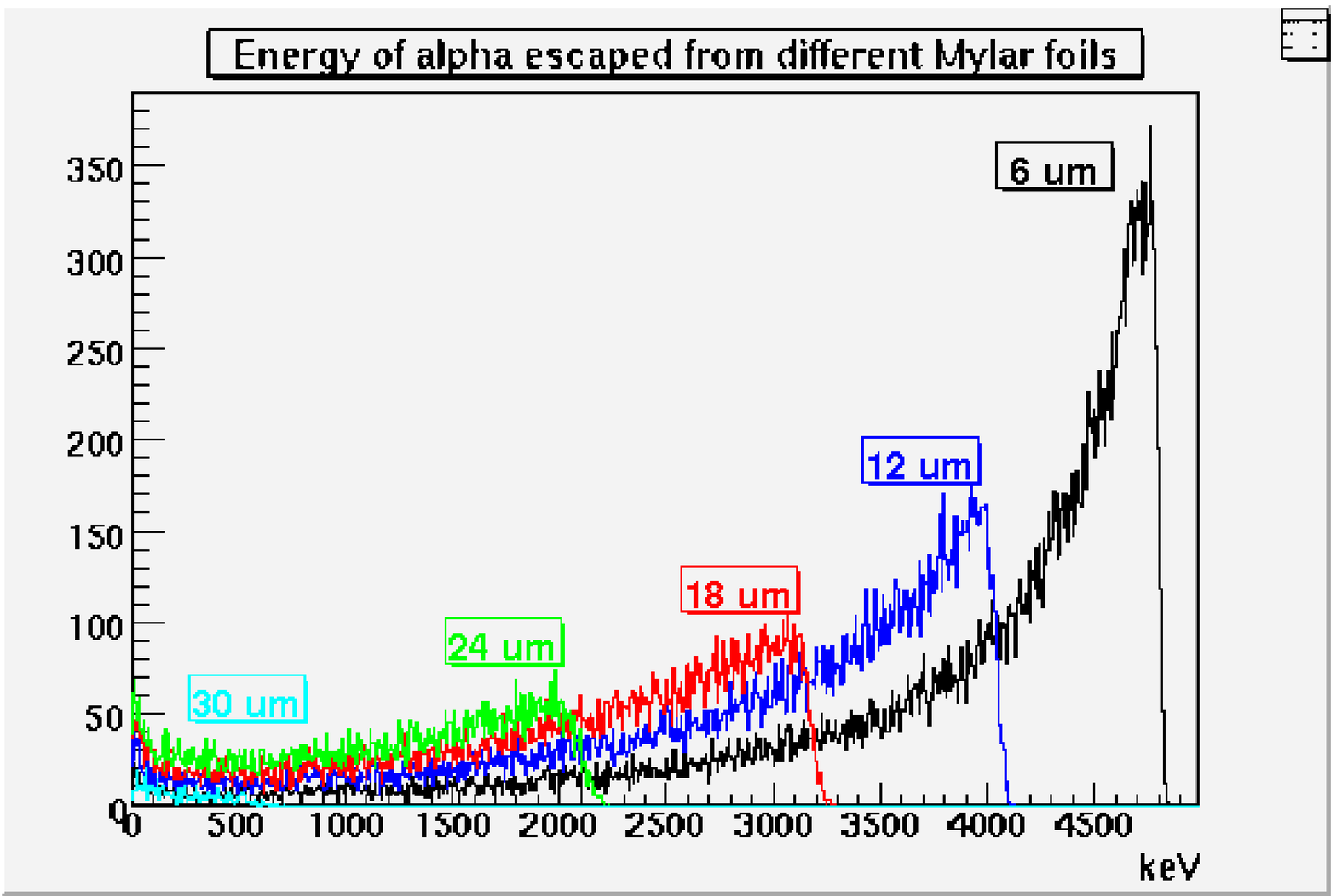}
  \includegraphics[height=.17\textheight]{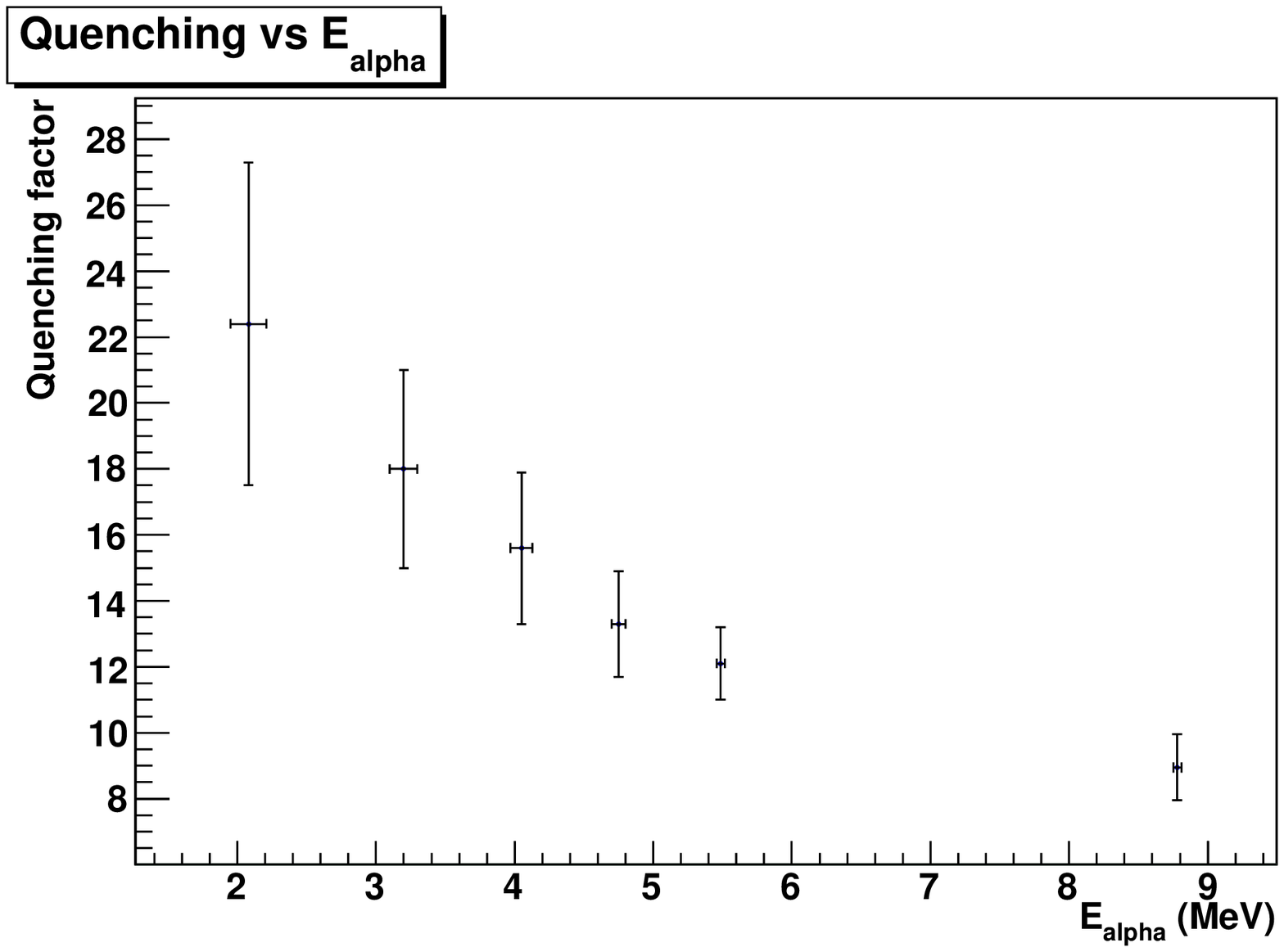}
  \vspace*{0,2cm}
\caption{Measurement scheme - Energy spectrum of $\alpha$ escaping the foil for differents mylar's thickness (6~$\mu$m foils) - Measured quenching factors (the value at 8~MeV comes from a NEMO~3 measurement).}
  \label{fig:alphas}
\end{figure}

The quenching factor, as a function of the energy of the $\alpha$ particles, has been measured for the plastic scintillators used in BiPo prototype with a dedicated test bench. $\alpha$ particles of 5.5~MeV have been produced by an $^{241}$Am source. Theirs energies have been reduced by pilling up 6~$\mu$m mylar foils between the source and the scintillator. A GEANT~4 simulation of $\alpha$ particles emitted by $^{241}$Am and crossing several foils of mylar has been done in order to determine the expected $\alpha$ energy spectrum (figure~\ref{fig:alphas}). Then the quenching factor has been measured by comparing the observed energy with the one of 1~MeV electrons from a $^{207}$Bi source. The result is presented in figure~\ref{fig:alphas}. The quenching factor is about 25 for a 1~MeV $\alpha$. It means that the expected energy threshold for electron detection of 40~keV would correspond to an energy threshold for $\alpha$ detection of 1~MeV.
\vspace*{0,3cm}

\subsection{Efficiency and sensitivity}

The efficiency of the BiPo detector depends on the capacity for an $\alpha$ to escape the foil. It is thus function of the thickness of the foil to be measured and the energy threshold of the scintillator.

\begin{figure}[hhh]
\centering
  \includegraphics[height=.25\textheight]{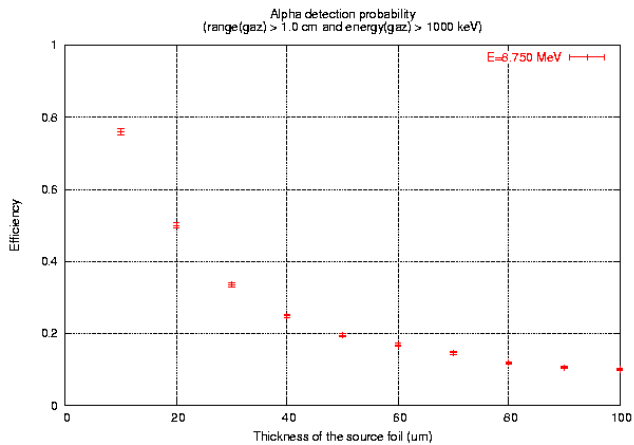}
  \caption{Efficiency to detect an alpha escaping the foil with energy greater than 1~MeV.}
  \label{fig:alphasefficiency}
\end{figure}

A GEANT4 Monte Carlo of 8.75~MeV $\alpha$ from $^{212}$Po decay, emitted randomly in the volume of a $^{82}$Se foil has been done. Figure~\ref{fig:alphasefficiency} shows the probability of these $\alpha$ to escape with an energy greater than 1 MeV (40 keV energy threshold for electron detection) for different thickness of foil. For a thickness of 40~mg/cm$^{2}$ (in the case of SuperNEMO foils), the efficiency is 25\%. This calculation is done with the hypothesis of bulk contamination. The efficiency to tag surface contamination would be obviously much larger.

The total efficiency of the full BiPo detector without tracking device has been calculated with a preliminary GEANT4 Monte Carlo. Assuming a thickness of the foil of 40~mg/cm$^{2}$, an uniform bulk contamination of $^{212}$Bi in the foil and an energy threshold of 40~keV for electrons (1~MeV for alpha), the total efficiency is 6.5\%. With a tracking device, the efficiency dropped to 2.5\%. This calculation takes into account rejection of the BiPo events in the case of a backscattering of the electron on one scintillator with a deposited energy above the threshold before hitting the second scintillator. Such events with two scintillators in time are rejected as candidate of $^{212}$Bi contamination in the scintillators (see below).

With a surface of the detection of 10~m$^{2}$ (corresponding to 5~kg of 40~mg/cm$^{2}$ foils to be measured), and with 1 month of measurement, a level of background of 1 count per month would correspond to a sensitivity in $^{208}$Tl activity of 2~$\mu$Bq/kg (1$\mu$Bq/m$^{2}$). In case of the design with a tracking device, the sensitivity would be 6~$\mu$Bq/kg (3$\mu$Bq/m$^{2}$).

\section{Background}

\subsection{Origin of the background}

The first limitation of the BiPo detector is a random coincidence of two scintillation hits within the 1~$\mu$s delay time window. Single counting rate is dominated by Compton electron due to external $\gamma$. It will be strongly reduced either by the tracking detector or by the external $\gamma$ tagger depending on the design of the BiPo detector. Thickness of scintillators must be also as thin as possible in order to reduce Compton electrons. The use of ultra thin scintillating fiber would also reduce strongly the random coincidence. One of the objective of prototypes is to demonstrate that the random coincidence will be smaller than 1~count per month for a surface of measurement of 10~m$^2$.

\begin{figure}[htb]
\centering
  \includegraphics[scale=0.38]{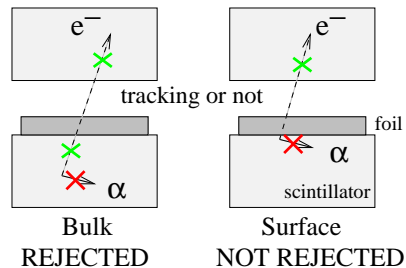}
  \caption{BiPo background from the scintillators : bulk and surface contamination. Crosses represent hits in scintillators.}
  \label{fig:bckg}
\end{figure}

The main source of background mimicking a BiPo (electron, delayed alpha) event is a $^{212}$Bi surface contamination on the entrance surface of the scintillators block in front of the foil as it is shown in figure~\ref{fig:bckg}. If the deepness of the contamination is small (typically less than 100~$\mu$m\footnote{The averaged deposited energy of an electron from the $^{212}$Bi beta decay calculated with a GEANT4 Monte Carlo is about 50~keV in 100~$\mu$m of plastic scintillator.\label{foot:deepness}}), the electron from the $^{212}$Bi beta decay will escape the first scintillator and hit the second one without depositing enough energy to trigger the first one. It will appear exactly like a BiPo event emitted from the foil.
 
However a bulk $^{212}$Bi contamination inside the block of the scintillators is not a source of background because the emitted electron will fire the scintillators block before escaping and hitting the second one. The two fired scintillators block will be in time and this background event will be rejected.

The level of surface purity in $^{212}$Bi of the scintillators required for the BiPo detector cannot be measured with HPGe germanium detectors. So in order to validate the surface radiopurity of the scintillators and also to validate the technique and the level of random coincidence, prototypes are under construction. 

\subsection{Preliminary measurements with NEMO~3 detector}

As the technology of the first design is the same as NEMO~3 detector, we used it to do some measurements of the $^{212}$Bi contaminations of the plastic scintillators (bulk and surface). To test the $^{212}$Bi contamination of the surface an analysis was done with NEMO~3 data. The analysis channel was a crossing electron in the wire chamber without counter triggered at the beginning of the track and a delayed hit in this scintillator. The decay spectrum was well fitted corresponding to the 300~ns half-life with a measured activity of 150~$\mu$Bq/m$^{2}$. This value is to much for BiPo detector because it corresponds to 400~$\mu$Bq/kg for 40~mg/cm$^{2}$ foils. This result isn't concluding because in NEMO~3 the scintillators are wrapped with aluminized mylar for light reflectivity. Our hypothesis is that this contamination comes from the standard aluminum deposited on this mylar. A BiPo prototype is still needed.

The $^{212}$Bi bulk contamination of the NEMO~3 scintillators was measured directly on a NEMO~3 counter with an oscilloscope looking for a PMT signal with a delayed hit within 1~$\mu$s. After 3 days no event has been observed. It corresponds to an upper limit on the $^{212}$Bi bulk contamination of the scintillators of A($^{212}$Bi)~$<$~3~$\mu$Bq/kg, in agreement with the BiPo requirements. This preliminary test will be confirmed with the BiPo-1 prototype (see below).

\section{Prototypes of the BiPo detector}

Two prototypes BiPo-1 and BiPo-2 are under construction. The main goal of these prototypes is to measure the level of background: (i) the level of random coincidence; (ii) the surface radiopurity of the scintillators.

These two prototypes will be installed in the new Canfranc Underground Laboratory in a shielding test facility. It consists of a large tight mechanical structure with 10~cm of low active lead (20~Bq/kg) and a inner layer of 4~cm of pure iron (in order to suppress bremsstrahlung from the lead shielding). The inner volume of the shielding is 140$\times$140$\times$100~cm$^3$ and will be flushed with radon free air.
 
\subsection{Prototype BiPo-1}

\begin{figure}[htb]
\centering
  \includegraphics[scale=0.43]{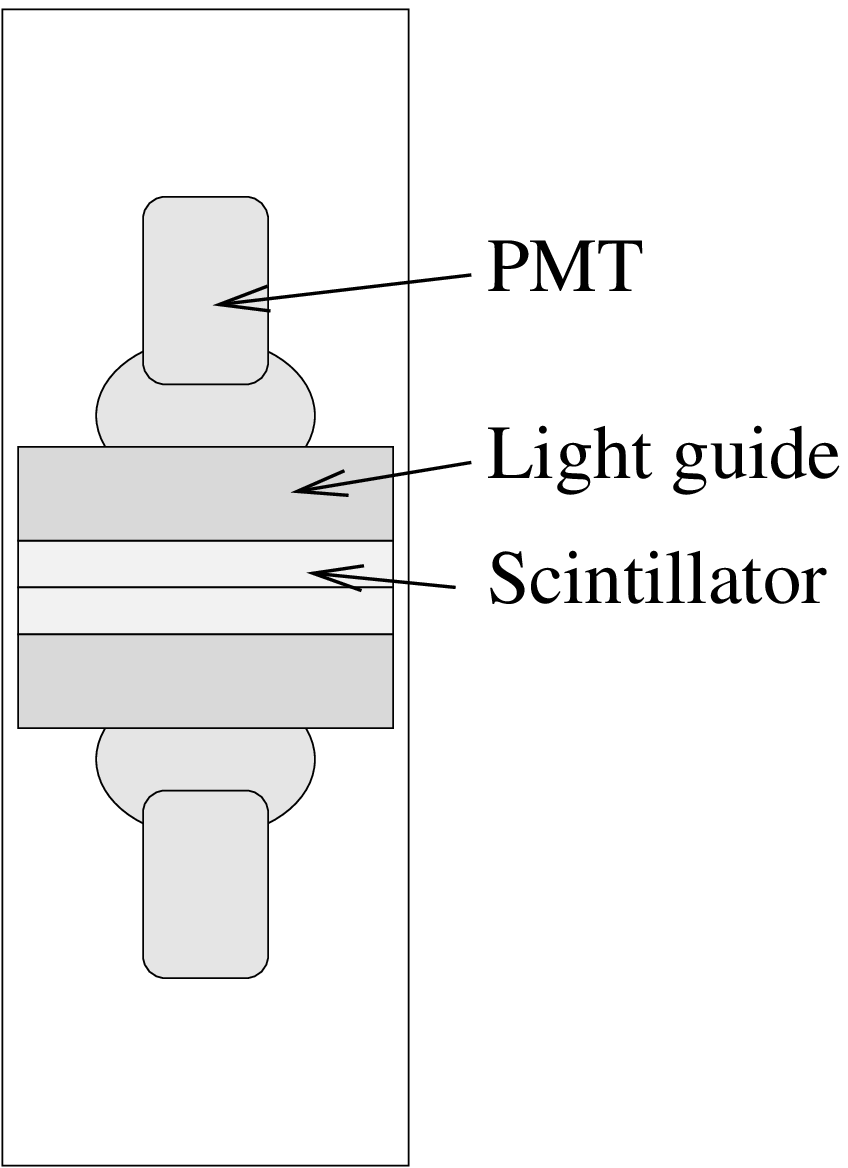}
  \hspace{1.5cm}
  \includegraphics[scale=0.38]{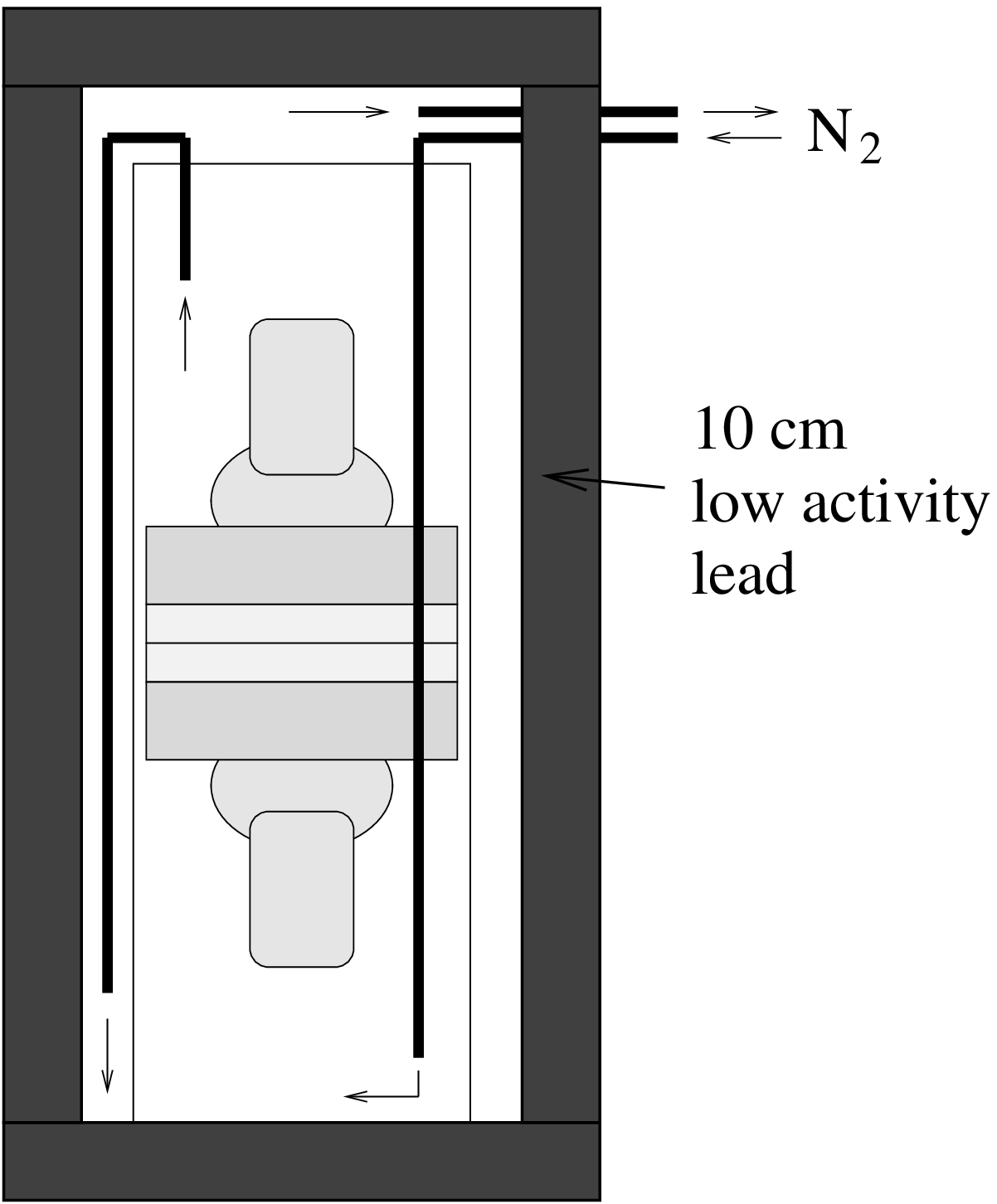}
  \caption{Scheme of a capsule of the BiPo-1 prototype and the first capsule with its own shielding in Canfranc.}
  \label{fig:capsule}
\end{figure}

The first BiPo-1 prototype (figure~\ref{fig:capsule}), consists of 20 low radioactive capsules made with Plexiglas or carbon fibers containing two organic plastic scintillators blocks face-to-face coupled with PMMA optical light guide to 5'' low radioactive PMTs (Hamamatsu R6594, 0.7~Bq/PMT total activity). The size of the scintillator blocks are 20$\times$20$\times$1~cm$^{3}$. Their entrance surface is covered with 200~nm of ultra-pure aluminum in order to avoid any scintillation light crosstalk. The capsules are filled by pure nitrogen in order to suppress radon and thoron contamination. 

Surface radiopurity in $^{212}$Bi of the entrance surface of the scintillators is measured by looking for a hit in one PMT and a delayed hit in the second PMT within 2~$\mu$s ($\sim$7$\times$T$_{1/2}$~($^{212}$Po)) without any hit in time (bulk contamination) or short delay in within 10~ns (double Compton from external $\gamma$). The large time window allows to distinguish BiPo candidates (up to 1~$\mu$s) from random coincidence (flat distribution). Acquisition of the full prototype will be done with the MATACQ acquisition board, a very high dynamic range and high sampling rate VME digitizing ADC board. It is a 4 channel boards with an amplitude resolution of 12~bits, a time range of 2.5~$\mu$s, an amplitude range of 1~V and a sampling rate of 2~Gs/s.

A first capsule has been installed in Canfranc Underground Laboratory in October, 17$^{th}$ shielded with only 10~cm of low activity lead and flushed with N$_{2}$ gas. The single counting rate was 0.3~Hz with an energy threshold of 60 keV. It has been reduced to about 0.1~Hz by adding only 1~cm of pure iron between the capsule and the lead, demonstrating that counting rate is dominated by bremsstrahlung from lead activity. A temporary acquisition with LECROY WAVERUNNER digital oscilloscope (2.5~GS/s and 8~bit amplitude resolution) has been used. After 10.7 days of measurement no BiPo event has been observed with the 1~$\mu$s delay window and one event has been observed with a 1.4~$\mu$s delay, in agreement with 0.33 count expected from random coincidence. It corresponds to an upper limit on the surface radiopurity of the scintillators of A($^{208}$Tl)~$<$~60~$\mu$Bq/m$^2$ (90\% C.~L.).

In September 2007, 20 capsules will be fully running in the shielding test facility in Canfranc. One month of measurement will allow to reach the sensitivity for the surface radiopurity of the scintillators of A($^{208}$Tl)~$<$~1~$\mu$Bq/m$^2$ (90\% C.~L.), level required for the final BiPo detector. This first BiPo-1 prototype could be used as a first BiPo detector with a capacity to measure a surface of 0.8~m$^2$ of double beta foils with a sensitivity of A($^{208}$Tl)~$<$~20~$\mu$Bq/m$^2$ (90\% C.~L.).

\subsection{Prototype BiPo-2}

The second BiPo-2 prototype consists of two face-to-face large scintillator plates (75$\times$75~cm$^2$ and 1~cm of thickness). The scintillating plates are polished without any wrapping in order to collect the scintillation light on the lateral sides by total internal reflectivity. The optical readout is done for each plate on two opposite sides with low radioactive PMTs coupled with optical light guides. The position of the emission of scintillation light is reconstructed by the barycentre of the amount of light of each PMT. A gain survey setup done with a LED sent with optical fibers on each PMT allows to control the stability of the PMTs gains. The plates are surrounded by a $\gamma$ tagger.

This prototype will be installed by the fall 2007 in the shielding test facility in Canfranc with the same ADC boards than for prototype BiPo-1. The objectives of this prototype are: (i) to validate the spatial resolution of this device, (ii) to measure the level of random coincidence depending on the spatial resolution and $\gamma$ tagger efficiency, (iii) the level of surface radiopurity of the scintillators plates. With two months of measurement, and if the random coincidence is low enough, the expected sensitivity of the prototype BiPo-2 is A($^{208}$Tl)~$<$~15~$\mu$Bq/m$^2$\linebreak (90\% C.~L.).

\section{Conclusion}

The BiPo detector with its two possible designs, is mainly purposed to the measurement of the SuperNEMO double beta source foils with a sensitivity of 2~$\mu$Bq/kg in $^{208}$Tl and 10~$\mu$Bq/kg in $^{214}$Bi. The R\&D phase started this year and will be going on for three years with the development of two prototypes BiPo-1 and BiPo-2. The main goal of these prototypes is to measure the level of backgrounds dominated by random coincidences and surface radiopurity of the scintillators. In October, 17$^{th}$ a first capsule of the prototype BiPo-1 as been installed in Canfranc Underground Laboratory. With 10 days of data, a preliminary upper limit on the surface radiopurity of the scintillators of A($^{208}$Tl)~$<$~60~$\mu$Bq/m$^2$ (90\% C.~L.) has been obtained. In September 2007, 20 capsules will be fully running. One month of measurement will allow to reach the sensitivity for the surface radiopurity of the scintillators of A($^{208}$Tl)~$<$~1~$\mu$Bq/m$^2$ (90\% C.~L.), level required for the final BiPo detector. By the end of 2007, the second design with scintillator plates will be tested for validation.

\end{document}